\documentclass[aps,12pt,nobibnotes,nofootinbib,prl,superscriptaddress,preprint]{revtex4-1}
\def\beq{\begin{eqnarray}}
\def\eeq{\end{eqnarray}}
\def\be{\begin{eqnarray}}
\def\ee{\end{eqnarray}}

\def\ba{\begin{eqnarray}}
\def\ea{\end{eqnarray}}
\def\beq{\begin{eqnarray}}
\def\eeq{\end{eqnarray}}

\def\mpl{M_{\rm Pl}}

\def\p{{\cal P}}

\def\L*{{\cal L}_*}
\def\L{\mathcal{L}}
\def\({\left(}
\def\){\right)}

\def\nn{\nonumber}
\def\p{\partial}
\def\mn{_{\mu \nu}}
\def\stu{St\"uckelberg }
\def\p{\partial}

\def\<{\langle}
\def\>{\rangle}

\usepackage{amsmath}
\usepackage{amsfonts}
\usepackage{verbatim}
\usepackage{setspace}
\usepackage{url}
\usepackage{color}
\usepackage{graphicx}

\newcommand{\pht}{\phantom}
\newcommand{\ud}{\mathrm{d}}
\newcommand{\uD}{\mathrm{D}}

\newcommand{\Comment}[1]{{}}
\definecolor{MyDarkBlue}{rgb}{0.15,0.15,0.45}
\usepackage[linktocpage=true]{hyperref}
\hypersetup{
colorlinks=true,
citecolor=MyDarkBlue,
linkcolor=MyDarkBlue,
urlcolor=MyDarkBlue,
}

\begin{document}

\preprint{UCSD/PTH 13-11 }
\preprint{NYU-TH-07/15/2013}

\title{On the Potential for General Relativity and its Geometry}


\author{Gregory Gabadadze}
\email[Electronic address: ]{gg32@nyu.edu}
\affiliation{Center for Cosmology and Particle Physics, Department of Physics,
New York University, New York, NY, 10003, USA}
\author{Kurt Hinterbichler}
\email[Electronic address: ]{khinterbichler@perimeterinstitute.ca}
\affiliation{Perimeter Institute for Theoretical Physics,
31 Caroline St. N, Waterloo, Ontario, Canada}
\author{David Pirtskhalava}
\email[Electronic address: ]{pirtskhalava@physics.ucsd.edu}
\affiliation{Department of Physics, University of California, San Diego, La Jolla, CA 92093}
\author{Yanwen Shang}
\email[Electronic address: ]{yshang@perimeterinstitute.ca}
\affiliation{Perimeter Institute for Theoretical Physics,
31 Caroline St. N, Waterloo, Ontario, Canada \vspace{1cm}}



\begin{abstract}

The unique ghost-free mass and nonlinear potential terms for general relativity
are presented in a diffeomorphism  and local Lorentz  invariant  vierbein formalism.
This construction  requires  an additional  two-index 
St\"uckelberg  field,  beyond the four scalar fields used in the  metric formulation, and unveils a 
new local SL(4) symmetry group of the  mass and potential terms, not shared by the Einstein-Hilbert term.  
The new field is auxiliary but transforms as a vector under two different Lorentz  groups, one of 
them the  group of local Lorentz transformations, the other an additional global group.  
This formulation enables a geometric interpretation of the 
mass and potential terms for gravity in terms of certain volume forms.
Furthermore, we find that the decoupling limit is much simpler to extract in this approach;  
 in particular, we are  able to derive expressions for the interactions of the vector modes.   
 We also note that it is possible to extend the theory by  promoting the two-index auxiliary field into a 
Nambu-Goldstone boson nonlinearly realizing a certain  space-time  symmetry,
and show how it is ``eaten up"  by the antisymmetric part of the  vierbein.

\end{abstract}


\maketitle

\section{1. Introduction and Summary}

Einstein's gravity is the theory that describes the two degrees of freedom of the massless helicity-2 
representation of the Poincar\'e group, and their two derivative self-interactions.
One may ask whether it is possible to alter the interactions of the graviton beyond those dictated by the 
Einstein - Hilbert (EH) action.  At the lowest, zero-derivative level, such a deformation would correspond to adding a potential for the metric perturbation.  An obvious example is the potential described by the cosmological constant (CC) term, ${\cal L}_0\sim \sqrt{-g}\Lambda$. This changes neither the number of propagating degrees of freedom of general relativity (GR), nor the consistency of the theory, but necessarily 
alters the background spacetime. 

The CC is the only such term -- other potentials  
inevitably change the number of degrees of freedom. 
The Fierz-Pauli  term \cite{Fierz:1939ix} is the unique consistent 
quadratic potential that gives rise to 5 degrees of freedom, as required by  
the massive spin-2  representation of the Poincar\'e group.   
Adding a  generic potential to  the EH action however leads to the loss of all 
four Hamiltonian constraints of GR, and thus a total of six  
propagating degrees of freedom, one of which is necessarily a ghost 
\cite{Boulware:1973my}. 

Nevertheless, there exists a special class of  mass and potential terms (the often-called dRGT terms \cite{deRham:2010ik,deRham:2010kj}, see \cite{Hinterbichler:2011tt} for a review) that make the graviton massive, while
retaining one of the four Hamiltonian constraints.  This remaining constraint projects out the ghostly sixth degree of freedom \cite {Hassan:2011hr,Hassan:2011ea},  see also  
\cite{Mirbabayi:2011aa,Hinterbichler:2012cn,Deffayet:2012nr}.

In addition to the CC term,  the dRGT construction allows for $3$ free parameters.  One combination is  
the graviton mass, $m$,  and the other two independent combinations, $\alpha_3$ and $\alpha_4$,  set the strength 
of the nonlinear potential.  The theory can be formulated by using four 
spurious diffeomorphism scalars, $\phi^{\bar a}$ --
first introduced  in an earlier proposal for massive gravity \cite {Siegel:1993sk} -- to 
allow for a  manifestly diffeomorphism-invariant  description. 
Adopting these four scalars, and following \cite {deRham:2010kj}, one can define a matrix with components
$\mathcal{K}^\mu_{~\nu}=\delta^\mu_\nu-\sqrt{g^{\mu\alpha}\p_\alpha\phi^{\bar a}
\p_\nu\phi^{\bar b}\eta_{\bar a\bar b}}$~, that can be used to build 
invariants supplementing  the EH action by the graviton mass 
as well as  zero-derivative interactions that guarantee 5 
degrees of freedom on an arbitrary background. One such term is given by \cite{deRham:2010kj} 
\beq
\label{u2}
{\cal L}_2\sim \frac{\mpl^2 m^2}{2}~\sqrt{-g}\varepsilon_{\mu_1\mu_2\bullet\bullet}
\varepsilon^{\nu_1\nu_2\bullet\bullet}\mathcal{K}^{\mu_1}_{~\nu_1}\mathcal{K}^{\mu_2}_{~\nu_2}~.
\eeq
The remaining two possible terms ${\cal L}_{3,4}$~, cubic and quartic in $\mathcal{K}$ respectively, 
can be obtained by the higher order generalization of \eqref{u2} \footnote{The $\epsilon$'s here are the 
epsilon symbols, with no factors of $\sqrt {-g}$. Moreover, the linear term 
${\cal L}_1\sim\sqrt{-g}\varepsilon\varepsilon\mathcal{K}$ can be expressed -- up to a total derivative --
through a linear combination of ${\cal L}_{2,3,4}$  and the CC.}, 
\beq
\label{u34}
{\cal L}_3 \sim \alpha_3 \mpl^2 m^2~\sqrt{-g}\varepsilon_{\mu_1\mu_2\mu_3\bullet}\varepsilon^{\nu_1\nu_2\nu_3\bullet}
\mathcal{K}^{\mu_1}_{~\nu_1}\mathcal{K}^{\mu_2}_{~\nu_2}\mathcal{K}^{\mu_3}_{~\nu_3}, \\
{\cal L}_4 \sim \alpha_4 \mpl^2 m^2~\sqrt{-g}\varepsilon_{\mu_1\mu_2\mu_3\mu_4}\varepsilon^{\nu_1\nu_2\nu_3\nu_4}
\mathcal{K}^{\mu_1}_{~\nu_1}\mathcal{K}^{\mu_2}_{~\nu_2}\mathcal{K}^{\mu_3}_{~\nu_3}\mathcal{K}^{\mu_4}_{~\nu_4}. 
\eeq

In addition to being invariant under the global Poincar\'e subgroup,  
$ISO(3,1)_{\text{GCT}}$ , of the group of general coordinate transformations (GCT), 
the theory is  invariant under an additional, global internal 
Poincar\'e group, $ISO(3,1)_{\text{INT}}$,  realized on the ``flavor" indices of 
the scalars, as first pointed out by Siegel in an earlier context \cite {Siegel:1993sk}
\be 
\phi^{\bar a}\to L^{\bar a}_{~\bar b}\phi^{\bar{b}}+c^{\bar b}. 
\label{Siegel}
\ee
Generation of the graviton mass occurs in the phase defined by the vacuum expectation 
value (VEV) of the order parameter $\langle \p_\mu\phi^{\bar a}\rangle=\delta^{\bar a}_{\mu} $. 
This results in the spontaneous symmetry breaking pattern of the global symmetry group
\be 
ISO(3,1)_{\text{GCT}}\times ISO(3,1)_{\text{INT}}\to ISO(3,1)_{\text{ST}}.
\ee
The unbroken $ISO(3,1)_{\text{ST}}$ group guarantees that the resulting theory is invariant under the 
ordinary spacetime  (ST) Poincar\'e transformations. Three of the four auxiliary scalars $\phi^{\bar a}$ 
are ``eaten" by the graviton to form a massive spin-2 representation of the latter group, while the fourth, potentially ghostly scalar is made non-dynamical by the single remaining Hamiltonian constraint of massive GR, originating from the specific structure of the dRGT terms ${\cal L}_{2,3,4}$.

The dRGT theory  gets rid of the sixth  ghostly mode, 
and also guarantees that the remaining 5 are unitary degrees of freedom  
at low energies  and on nearly-Minkowski  backgrounds (i.e., the backgrounds  with 
typical curvature smaller than the graviton mass square).
However,  the theory does not guarantee that  for 
more general backgrounds  the 5  physical modes are healthy. 
In fact, some of their kinetic terms may change signs around  
 certain  cosmological backgrounds. Moreover,  for a large 
region of the $\alpha_2,\alpha_3$ parameter space,   the potential is known to
violate  the null energy condition and one often  gets kinetic and gradient terms 
that give rise to superluminal group and phase velocities.
Most of the above issues  stem from one and the same source:
the dRGT theory is strongly coupled at the energy/momentum scale
$\Lambda_3 \equiv (M_{Pl} m^2)^{1/3}$  \cite{deRham:2010ik,
deRham:2010kj}.   As a result, a typical  curvature of order  $m^2$ produces 
order 1 corrections to the kinetic terms for fluctuations, often giving 
rise to vanishing or negative kinetic terms, or superluminal  group and 
phase velocities (for brief comments on the current state of affairs on all 
these issues,  see  Section 6).

  As for any strongly coupled theory, an extension 
above the scale $\Lambda_3$ is desirable\footnote {Using  the 
particle physics terminology,  dRGT is a theory 
with no ``radial mode",  i.e., the graviton gets a mass via the 
Anderson mechanism,  as opposed to the Higgs mechanism. 
What may be needed  is an extension to include putative  ``radial mode(s)" 
that would ensure weakly coupled behavior above $\Lambda_3$.}. 
However,  it is hard to think of such an extension since  the Lagrangian 
contains square roots of the longitudinal modes 
(represented  by the $\phi^{\bar a}$'s).  This inconvenience might 
be mitigated by using the vierbeins, which are square roots of the metric.  
The goal of the present work is to rewrite the theory in terms of  the vierbeins
in a GCT and local Lorentz transformation (LLT) invariant form.  The hope is that this form 
of the theory might make it easier to find a weakly coupled completion. 
Also, irrespectively of that, the vierbein formulation itself merits a 
separate consideration. 

A vierbein reformulation of the theory was given by one of us  and 
R. A. Rosen\footnote{Ref.  
\cite {Ali} has extracted the square root in dRGT using vierbeins,  however, we disagree 
with the main conclusion of that work on the Boulware-Deser  degree of freedom. 
See also Refs. \cite {Ali1,Ali2} for earlier interesting works on the vierbein formulation of bigravity.} \cite {Hinterbichler:2012cn}.  That work focused on a unitary gauge description, which 
for a single massive graviton is not GCT or LLT invariant.  In the present work, we give a GCT and LLT 
invariant action for a massive graviton.  

We find that such a formulation requires  a new  two-index  
St\"uckelberg  field, $\lambda^a_{~\bar a}$,  in addition 
to the four scalar fields $\phi^{\bar a}$ used in the  metric description. 
The new  field is auxiliary   and enters the action algebraically.  To recover dRGT,
this field  should transforms as a vector under two different Lorentz  groups, 
$ \lambda^a_{~\bar a} \to Q^a_{~b}(x) \lambda^b_{~\bar a} $,  and 
$ \lambda^a_{~\bar a} \to L^{\bar b}_{~\bar a} \lambda^a_{~\bar b} $,
where $Q(x)$ belongs to $SO(3,1)_{\rm LLT}$, while the constant matrix 
$L$  belongs to the  global group  $SO(3,1)_{\rm INT}$.  
Moreover,  we note that the mass and potential terms -- once 
written in the  GCT and LLT invariant form -- are amenable to an extension 
with $\lambda \in SL(4)$ and  unveil a new 
local symmetry w.r.t. simultaneous transformations,
$ e^{~a}_\mu \to Q^a_{~b}(x) e^{~b}_\mu$ and 
 $ \lambda^a_{~\bar a} \to Q^a_{~b}(x) \lambda^b_{~\bar a} $,
 where $Q(x) \in SL(4)$.   Thus, the enhanced symmetry group
 of the mass and potential terms,  $SL(4)\times G_{\rm GCT}$, is larger than 
 the  symmetry group of the EH action. This  observation suggests 
 an extension of the theory by additional fields (see Section 2, and  Section 3 for 
 a $GL(4)$ symmetric extension).  

As we will discuss  in Section 3, the vierbein formulation enables one to give 
a geometric interpretation to the  mass and potential terms --
they can be expressed in terms of certain volume forms.

There are other benefits as well:  we find that the decoupling limit is 
much simpler to extract in this approach. The original results of  
\cite{deRham:2010ik} can be obtained with significantly less effort. 
Moreover, it is straightforward to derive  closed--form expressions for the vector 
modes,  which have not been obtained in complete generality before.

We also note that the field $ \lambda^a_{~\bar a} \in SO(3,1)$ can be represented as 
$\lambda^a_{~\bar a} = {\rm exp} ( v^{a}_{~\bar a}/f)$, where $v$ is an antisymmetric field (once indices are lowered with $\eta$) and $f$ 
is some dimensionful constant. Then,  $v$ can be promoted  
into a dynamical Nambu-Goldstone field parametrizing a coset 
$\(SO(3,1)_{\text{GCT}}\times SO(3,1)_{\text{INT}}\)/SO(3,1)_{\rm Diag}$. 
We show that these six bosons are ``eaten up"  by the antisymmetric part of the  vierbein.
This extends ghost-free massive gravity to a theory where the six antisymmetric components 
of the vierbein  become  dynamical.

\section{2. Vierbein formulation}

The formulation of massive GR, as well as its extensions, is significantly simplified in the vierbein formalism \cite{Hinterbichler:2012cn}. Introducing the vierbein field $e^{\ a}_{\mu}$, $g_{\mu\nu}=e^{\ a}_{\mu}e^{\ b}_{\nu}\eta_{ab}$ with $\eta_{ab}=diag(-1,1,1,1)$, the cosmological constant term can be written as $d^4x {\cal L}_0\sim d^4x \sqrt{-g}\Lambda\sim \Lambda ~\varepsilon_{abcd}~ e^a\wedge e^b \wedge e^c \wedge e^d$~,
where the one form $e^a$ is defined as $e^a\equiv e^{\ a}_{\mu}\text{d}x^\mu$.  The ghost-free interactions of the vierbein perturbations can be represented in a similar fashion; \textit{e.g.} in the unitary gauge, one such term is given by 
\be
d^4x {\cal L}_2 \sim \mpl^2 m^2\varepsilon_{abcd}~
e^a \wedge e^b\wedge  (e^c-\mathbf{1}^c)\wedge
(e^d-\mathbf{1}^d)\nn ~, 
\ee
where $\mathbf{1}^a\equiv \delta^{~a}_\mu \text{d}x^\mu$ represents a unit vierbein. The two contributions ${\cal L}_{0,2}$ to the potential can be supplemented by the two other independent terms ${\cal L}_{3,4}$, involving respectively three and four powers of $(e-\mathbf{1})$, contracted with the $\varepsilon$ symbol in a similar fashion\footnote{As in the second-order case, the remaining possible term ${\cal L}_1$, linear in $(e-\mathbf{1})$, can be expressed as a combination of the rest of the terms.},
\beq
\label{u34v}
d^4 x {\cal L}_3\sim\mpl^2 m^2\varepsilon_{abcd}~
e^a \wedge(e^b-\mathbf{1}^b)\wedge  (e^c-\mathbf{1}^c)\wedge
(e^d-\mathbf{1}^d) ~, \\ 
d^4 x {\cal L}_4\sim\mpl^2 m^2\varepsilon_{abcd}~
(e^a-\mathbf{1}^a) \wedge(e^b-\mathbf{1}^b)\wedge  (e^c-\mathbf{1}^c)\wedge
(e^d-\mathbf{1}^d)\,.
\eeq

The above terms together with the Einstein-Hilbert term define an action for the $16$ variables in the vierbein which is neither GCT nor LLT invariant, 
whereas the metric formulation is an action for $10$ metric variables (plus four scalars in the St\"uckelberg formulation).  
Nevertheless, both formulations are dynamically equivalent.  Following \cite {Hinterbichler:2012cn}, we first show that the vierbein action is dynamically equivalent to the same action only with the additional constraint that the vierbein is symmetric 
(with respect to the Minkowski metric).  In matrix notation,
\be
\label{symmviel}
e\, \eta= \eta\, e^T\,.
\ee
We parametrize the general vierbein as a constrained vierbein $\hat e$ satisfying \eqref{symmviel}, times a Lorentz transformation, parametrized as the exponential of a matrix $\hat B$ (which is anti-symmetric with respect to $\eta$)\footnote{See \cite{Deffayet:2012zc} for more on this condition and its relation to the square roots of the metric formulation.}, 
\be 
\label{vierbeinlparam} 
e={\hat e} \, e^{{-\hat B}}\,,\ \ \ \eta\,{\hat B} = -{\hat B}^T\, \eta.
\ee
The $\hat B$'s do not enter the Einstein-Hilbert term, since this term is invariant under local Lorentz transformations.
Thus, the $6$ variables in $\hat B$ appear only in the mass and potential terms
(which in the metric formulation depend on  the inverse metric $g^{-1}$ through the matrix ${\cal K} = {\bf 1} - \sqrt {g^{-1}\partial\phi \partial\phi}$).
These fields therefore appear without derivatives -- they are auxiliary fields.  We now vary with respect to $\hat B$ and look at the equations of motion, in powers of $\hat B$.  The lowest order terms  contain no powers of $\hat B$ (other than the variation $\delta {\hat B}$).  Therefore, the only terms that appear at lowest order are  the ones containing traces of one power of $\delta {\hat B}$ along with powers of $\hat e^{-1}$.  Because $\hat e^{-1}$ is symmetric and $\delta {\hat B}$ antisymmetric, and because $\delta {\hat B}$ appears only linearly, the terms in the equations of motion linear in $\hat B$ all vanish.  This means that the equations of motion of $\hat B$ start linearly in $\hat B$, and are solved by $\hat B=0$.  Plugging this solution back into the action, we see that the action with unconstrained vierbeins is dynamically equivalent to the action with symmetric vierbeins.

To relate the potential with symmetric vierbeins to the potential in the metric formulation we use the matrix 
representation $g = e\, \eta \, e^T$,
\be
g^{-1} \eta= (e^{-1})^T \, \eta^{-1} \,e^{-1}\,\eta\,.
\ee
Using the parametrization \eqref{vierbeinlparam} and the symmetry property of $\hat e^{-1}$:
\be
\sqrt{g^{-1} \eta} =\( \hat e^{-1} \)^T\,.
\ee  
Thus in the unitary gauge, $\partial_\mu \phi^{\bar a} = \delta_\mu^{\bar a}$, we can write,
\be
 {\cal L}_{2,3,4}( \sqrt{g^{-1}\eta} ) = {\cal L}_{2,3,4}({\hat e}^{-1})  \,.
\ee

Due to the presence of the unit vierbein, in the form presented above the first order theory lacks invariance under both the GCT and LLT, characteristic of general relativity. Both of the symmetries however can be restored via corresponding \stu fields. For this, one introduces the auxiliary scalars $\phi^{\bar a}$, analogous to those of the metric description of  massive GR, as well as the ``link" field $\lambda^a_{~\bar a}$. 
The latter transforms  as a contravariant vector under the local Lorentz group,
$ \lambda^a_{~\bar a} \to Q^a_{~b}(x) \lambda^b_{~\bar a} $,  
where $Q(x)\in SO(3,1)_{\rm LLT}$,
and as a covariant vector under the global group  $SO(3,1)_{\rm INT}$, $ \lambda^a_{~\bar a} \to L^{\bar b}_{~\bar a} \lambda^a_{~\bar b} $.  
Using these fields, the mass and potential terms can be rewritten in a manifestly GCT $\times$ LLT-invariant 
form via the ``k-vierbein",  $k^{~a}_\mu\equiv e^{~a}_\mu-\lambda^a_{~\bar a}\p_\mu\phi^{\bar a}$, 
\ba
\label{l2}
\mathcal{L}_2&\sim \mpl^2 m^2~\varepsilon^{\mu\nu\alpha\beta}\varepsilon_{abcd}
~e^{~a}_\mu e^{~b}_\nu k^{~c}_\alpha k^{~d}_\beta~ , 
\\ \label{l3}
\mathcal{L}_3&\sim \alpha_3\mpl^2 m^2~\varepsilon^{\mu\nu\alpha\beta}\varepsilon_{abcd}
~e^{~a}_\mu k^{~b}_\nu k^{~c}_\alpha k^{~d}_\beta~ , \\ \label{l4}
\mathcal{L}_4&\sim \alpha_4 \mpl^2 m^2~\varepsilon^{\mu\nu\alpha\beta}\varepsilon_{abcd}
~k^{~a}_\mu k^{~b}_\nu k^{~c}_\alpha k^{~d}_\beta~ \,.
\ea
(As before, the $\epsilon$'s here are the epsilon symbols, i.e. there are no factors of $\sqrt{-g}$.)
In the unitary gauge defined by $\lambda^{\bar a}_{~a}=\delta^{\bar a}_a$ and $\p_\mu\phi^{\bar a}=\delta^{\bar a}_\mu$,  one recovers the ${\cal L}_{2,3,4}$  of \eqref{u2}. Away from this gauge, the theory acquires invariance under GCT, as well as under LLT, realized on the vierbein and the link fields as follows
\beq
e^{~a}_\mu\to Q(x)^a_{~b} e^{~b}_\mu, \quad \lambda^a_{~\bar a}\to Q(x)^a_{~b} \lambda^b_{~\bar a}~.
\label{SO31}
\eeq

The transformations \eqref {SO31} with $Q(x) \in SO(3,1)_{\rm LLT} $ represent 
a  symmetry of the entire action, the potentials \eqref {l2} -\eqref {l4}  and the Einstein-Hibert term.
However,  the potential terms themselves, \eqref {l2}-\eqref {l4},   without the EH term,  
can have  a  larger symmetry. To see this, we first note that these potentials are invariant under the formal field redefinition \eqref {SO31} with $ Q(x)\in SL(4)$. 
Now, we  defined $\lambda$ to be a $SO(3,1)$ matrix  and therefore, such transformations 
with $SL(4)$ matrices would take them  outside of $SO(3,1)$.  This observation suggests
that in the theory where the EH term is absent, the $\lambda$  can be promoted to a 
$SL(4)$ valued field. The resulting terms, \eqref {l2}-\eqref {l4}, will have a local $SL(4)$ symmetry,
in addition to being invariant under GCT's. This extended local $SL(4)$  symmetry is the defining 
property of the mass and potential terms.

However, the EH term does not respect the $SL(4)$. Therefore,  there are two ways
to combine the EH term with the potentials \eqref {l2}-\eqref {l4}: (1) To define a theory where 
$\lambda$ is an $SO(3,1)$ valued field; (2) Alternatively, to define a theory with  
$\lambda \in SL(4)$. In this paper we chose the former case 
because that is the theory  of a single massive graviton.
The latter choice gives a theory with $9 = {\rm dim ~SL(4) } - {\rm dim ~SO(3,1)}$ 
additional fields, and might be an interesting model  to look at in the future.

Thus, for $\lambda \in SO(3,1)$,  in the unitary gauge, 
$\lambda = {\bf 1}$, with  $\phi^{\bar a}$ kept unfixed, one 
recovers the GCT-invariant but LLT 
non-invariant  formulation of massive GR \footnote{In this gauge  $k^{~a}_\mu = e^{~a}_\mu - \delta^a_{\bar a}
\partial_\mu \phi^{\bar a}$,  
and  the global Siegel's $ ISO(3,1)_{\rm INT} $  symmetry \eqref {Siegel}
gets  enhanced to a symmetry w.r.t. the global $ISL(4)$ transformations 
of the $\phi^{\bar a}$  fields,  if the vierbein is also transformed under the global $SL(4)$.   
The existence of this enhanced  global symmetry of the mass and potential terms 
had been pointed out  by W. Siegel in a private communication in Spring 2012,  and has recently   made us  
realize that in the LLT invariant theory  the potentials \eqref{l2}-\eqref{l4}  can be 
promoted to the local $SL(4)$ symmetric form.}.
The relevant symmetry breaking pattern, corresponding to this case, 
\beq
\label{sb}
SO(3,1)_{\text{INT}}\times SO(3,1)_{\text{LLT}} \to SO(3,1)_{\text{DIAG}}~,
\eeq
involves six broken generators, while the remaining six correspond to the diagonal part of LLT and internal Lorentz groups. The equation of motion for $\lambda$, evaluated in the unitary gauge, gives precisely the constraint \eqref{symmviel}, needed for the theory  to reduce to massive GR. 

As already remarked above, it is useful to represent the  vierbein as  $e^{~a}_\mu = {\rm exp} ({\hat B}^a_{~b}) {\hat e}^{~b}_\mu$ and 
the $\lambda$ field as $\lambda^a_{~\bar a} = {\rm exp} ( v^{a}_{~\bar a}/f)$,
where both $B$ and $v$ are  antisymmetric fields (once indices are lowered by $\eta$).   
Under  LLT, both of these fields shift by a 
coordinate dependent gauge function, so one or the other of them may be gauged away, but 
not both.  One linear combination of $\hat B$ and $v$ is invariant under LLT. 
This combination has no kinetic term 
in  our construction, and it is algebraically determined by classical equations 
of motion guaranteeing, by the same arguments given earlier.  Only the five helicities of the graviton are propagating degrees of freedom in the theory.  

An interesting alternative is to give dynamics to the gauge-invariant combination  
by regarding it as a Nambu-Goldstone field parametrizing 
the coset  corresponding to the symmetry breaking 
pattern  \eqref{sb}. The kinetic term  for this field also breaks the local  
$SL(4)$ of the potential down to the group  of LLT's.  We will discuss  this possibility in 
Section 5. Before then we will stay in the framework 
of massive gravity  and the gauge invariant part of the $\lambda$ field will be regarded as 
non-dynamical.

\section{3. Geometric interpretation and generalizations}

In this Section, we will give this formulation of the theory a geometric interpretation.
Let us consider two manifolds of the same dimension\footnote{We assume both manifolds to be dimension
$4$ for brevity in the current discussion, but this can be straightforwardly generalized, including to manifolds of different dimension, giving theories with new scalar degrees of freedom, along the lines of \cite{Gabadadze:2012tr}.},
and a smooth mapping between them
${\phi}:\mathcal M\rightarrow E$. When  a set of coordinates is given, 
the mapping $\phi$ consists of $4$ smooth functions which we denote by 
$\phi^a(x)$.  (We ignore any possible 
topological obstructions at the moment. Such a smooth mapping always exists locally within certain patches of both 
$\mathcal M$ and $E$.)

We denote, at each point of $x\in\mathcal M$ and $\phi(x)\in E$, the
cotangent spaces $T^\ast_{\mathcal M}(x)$ and $T^\ast_E(\phi)$ respectively.
A set of vierbeins $e^a=\ud x^\mu e^{~a}_{\mu}$ is defined for
every $T^\ast_{\mathcal M}(x)$ which endow $\mathcal M$
with a metric $g_{\mu\nu}\equiv e^{~a}_{\mu} e^{~b}_{\nu}\eta_{ab}$.
Usually, for the mapping $\phi$ between $\mathcal M$ and $E$ to be compatible
with their Riemannian structures, one must assume that
the metric on $\mathcal M$ coincides with the metric pulled back from $E$ through the functions $\phi^a(x)$, i.e. both manifolds 
share identical Riemannian geometries and the mapping $\phi$ represents 
nothing other than a simple coordinate transformation. Physically, the two 
are indistinguishable. 

If, on the other hand, we insist that the manifold $E$ should stay flat,
we may choose to define the vierbeins in each $T^\ast_E(\phi(x))$ as
$\theta^a=\ud\phi^a$. Together with the torsion-free condition, 
such a choice guarantees that the curvature tensor on $E$ vanishes.
But, such a construction of $\theta^a$ does not
respect the local Lorentz symmetry 
of $T^\ast_E(\phi(x))$, leaving only the global version intact\footnote{Formally, one may fix this 
by introducing another set of flat
spin-connections on $E$ and write $\theta^a=\uD \phi^a$ instead,
where $\uD$ is the covariant exterior derivative. 
It is not necessary for the current discussion and we choose not
to pursue this direction here.}.

Now that the two manifolds $\mathcal M$ and $E$ are endowed with totally different
Riemannian structures, there is no natural way to mix the cotangent vectors living
in $T^\ast_{\mathcal M}(x)$ and those living in $T^\ast_{E}(\phi(x))$. Indeed, if we
just write terms such as $e^a-\ud \phi^a$, they violate invariance  
w.r.t. the LLT's. The link fields  $\lambda^a_{\pht{a}\bar a}(x)$ are introduced
to remedy this.  Due to the specific transformation of 
$\lambda^a_{\pht{a}\bar a}$ under the two Lorentz groups, we are able to map the forms in one cotangent space
to the other and introduce mixing  via the ``k-vierbein" $e^a-\lambda^a_{\pht{a}\bar a}\ud\phi^{\bar a}$, 
where $\ud\phi^{\bar a} =  \partial_\mu \phi^{\bar a}dx^\mu$.  We write the mass and potential  in terms of the forms
\begin{equation}
\label{eq:mixing_terms}
\begin{split}
d^4 x {\cal L}_1\sim  \,&  \epsilon_{abcd}(e^a-\lambda^a_{\pht{a}\bar a}\ud\phi^{\bar a})\wedge
	e^b\wedge e^c\wedge e^d,\\
d^4 x {\cal L}_2 \sim  \,& \epsilon_{abcd}(e^a-\lambda^a_{\pht{a}\bar a}\ud\phi^{\bar a})\wedge
	(e^b-\lambda^b_{\pht{b}\bar b}\ud\phi^{\bar b})\wedge
	e^c\wedge e^d,\\
d^4 x {\cal L}_3 \sim  \,& \epsilon_{abcd}(e^a-\lambda^a_{\pht{a}\bar a}\ud\phi^{\bar a})\wedge
	(e^b-\lambda^b_{\pht{b}\bar b}\ud\phi^{\bar b})\wedge
	(e^c-\lambda^c_{\pht{c}\bar c}\ud\phi^{\bar c})\wedge e^d,\\
d^4 x {\cal L}_4 \sim \,& \epsilon_{abcd}(e^a-\lambda^a_{\pht{a}\bar a}\ud\phi^{\bar a})\wedge
	(e^b-\lambda^b_{\pht{b}\bar b}\ud\phi^{\bar b})\wedge
	(e^c-\lambda^c_{\pht{c}\bar c}\ud\phi^{\bar c})\wedge 
	(e^d-\lambda^d_{\pht{d}\bar d}\ud\phi^{\bar d}).\\
\end{split}
\end{equation}
As discussed in the  previous section, these expressions manifestly respect the local Lorentz 
symmetry on $\mathcal M$, defined by
\[
e^a\rightarrow  Q^a_{\pht{a}b} e^b\,\qquad
\lambda^a_{\pht{a}\bar b}\rightarrow Q^a_{\pht{a}c}\lambda^c_{\pht{c}\bar b}\,
\qquad \phi^{\bar a}\rightarrow \phi^{\bar a}\, ,
\]
at the cost of introducing the \stu fields $\lambda^a_{~\bar a}$. 

Notice that we can equally well 
write terms by multiplying $\lambda_{a}^{~\bar b}$ -- which
we define to be the inverse matrix of $\lambda^a_{\pht{a}\bar b}$ --  
onto  $e^a$ instead of $\ud\phi^{\bar a}$. So we could  write, as an example,
\[
d^4 x {\cal L}_2 \sim  \epsilon_{\bar a\bar b\bar c\bar d}(\lambda_a^{\pht{a}\bar a} e^a-\ud\phi^{\bar a})\wedge
	(\lambda_b^{\pht{a}\bar b} e^b   - \ud\phi^{\bar b})  \wedge 
	\lambda_c^{\pht{a}\bar c}e^c\wedge 
	\lambda_d^{\pht{a}\bar d}e^d\,.
\]
This formulation however is equivalent to the one in \eqref {eq:mixing_terms}.
In the  new form, the invariance under LLT is manifestly visible 
since  $\ud\phi^{\bar a}$ are invariant, and the LLT transformations 
of $e^a$ are simply compensated by the opposite rotation for $\lambda_a^{\pht{a}\bar a}$
so the combination $\lambda_a^{\pht{a}\bar a} e^a$ remains invariant automatically. 
Note that in this latter formulation  one can directly  extend $\lambda$ to a $GL(4)$-valued field,
and then have the mass and potential terms invariant under local $GL(4)$, instead of   
$SL(4)$ discussed  in Section 2. The $GL(4)$ invariant form can also be achieved  
in the original formulation, if the mass and potential terms \eqref {l2}-\eqref {l4} are multiplied 
by ${\rm det} (\lambda^{-1})$,  with $\lambda \in GL(4)$.

The terms in \eqref{eq:mixing_terms} are quite reminiscent of the CC
term in GR -- these terms strongly resemble 
some sort of volume forms.  In particular, one linear combination of the four terms  gives the CC term 
(up to a total derivative).  
If, for the moment, we imagine that the fields $\phi^{\bar a}$ 
are the embedding coordinates
of the manifold $\mathcal M$ into a higher dimensional
flat manifold (so that $\bar a$ takes the values of $1, 2,\dots, D$ where $D>4$), a term 
$\mathcal {L}\sim\lambda^a_{\pht{a}\bar a}\lambda^b_{\pht{b}\bar b}\lambda^c_{\pht{c}\bar c}
\lambda^d_{\pht{d}\bar d}~\ud\phi^{\bar a}\wedge\ud\phi^{\bar b}
\wedge\ud\phi^{\bar c}\wedge\ud\phi^{\bar d}~\varepsilon_{abcd}$, 
with a \emph{fixed}  matrix $\lambda^a_{\pht{a}\bar a}$ 
that projects the $D$-dimensional tangent vectors down to the tangent
space of $\mathcal M$, is the volume form for the surface $\mathcal M$ as embedded in $E$.

Here, in our formulation, there are two major differences. First of all,
we are dealing with the mixing terms among the vierbeins of two different
manifolds, $\mathcal M$ and $E$, with different geometries but an identical
dimensionality.  Secondly, we must
integrate w.r.t. all possible embeddings parameterized by $\lambda^a_{\pht{a}\bar b}$
to make a comparison between different volume forms meaningful.
Both differences complicate the geometrical identification of these
mixing terms.  However, for any fixed
$\lambda^a_{\pht{a}\bar b}$, each term in \eqref{eq:mixing_terms}
can be given a  geometric interpretation  in terms of a difference  
between certain volume forms of the two different manifolds.

Consider the simplest example, $d^4 x {\cal L}_1\sim(e^a-\lambda^a_{\pht{a}\bar a}\ud \phi^{\bar a})\wedge e^b\wedge e^c\wedge e^d~\epsilon_{abcd}$. Apart from the volume form of $\mathcal M$, it contains the term 
$\lambda^a_{\pht{a}\bar a}\ud \phi^{\bar a} \wedge e^b\wedge e^c\wedge e^d~\epsilon_{abcd}$.
If we choose the gauge $\lambda^a_{\pht{a}\bar a}=\delta^a_{~\bar a}$,  and focus only on the
term $(a, b, c, d)=(1, 2, 3, 4)$, we recognize this as the volume form
of $\mathcal M^3\times R$,  where $\mathcal M^3$ denotes  
a $3$-dimensional submanifold spanned out by the cotangent vectors 
$e^2$, $e^3$, and $e^4$, and $R$ denotes the ``flat dimension'' parameterized
by $\phi^1(x)$. So, ${\cal L}_1$ gives  a difference between
the two types of volume forms: the one  of $\mathcal M$ and another from 
those of $M^3\times R$,  with  $M^3$ now representing a 
3-dimensional submanifold of $\mathcal M$ spanned by any three of the four 
vierbeins $e^a$. Individually, each such term  depends on the arbitrary 
choice of $e^a$, $\phi^{\bar a}$, as well as the embedding matrix 
$\lambda^a_{\pht{a}\bar a}$,  but when all the indices are contracted 
and the fields are integrated over, we obtain a well-defined notion of a relative 
volume forms of the two manifolds.  

\begin{figure}[ht!]
\begin{center}
\includegraphics[width=0.688 \textwidth]{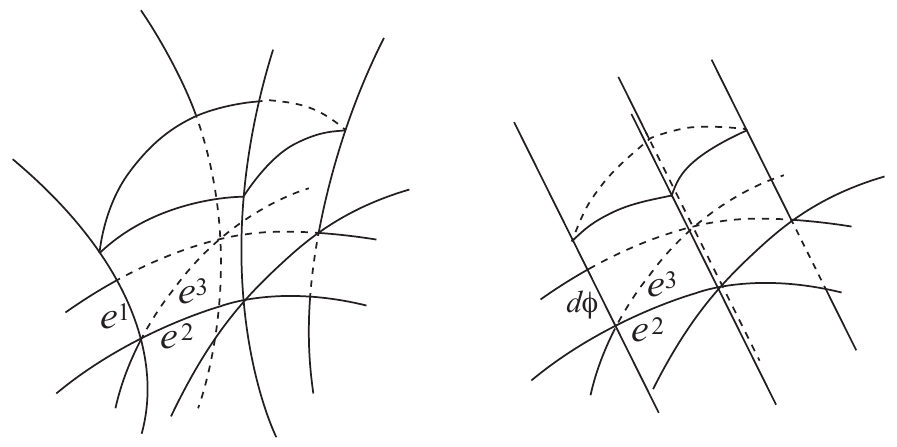}
\caption{\label{fig:volumes} Illustration of various volume forms, appearing in the graviton potential in massive GR.}
\end{center}
\end{figure}

Fig. \ref{fig:volumes} gives  an illustration to this.  The left figure represents
the original volume form of $\mathcal M$, with the 4-th dimension
suppressed, and the right one depicts the volume
form obtained  when the direction along that of $e^1$ is ``straightened''.
The difference between the two volume forms is $d^4x {\cal L}_1$.

Likewise, we may interpret terms 
$\lambda^a_{\pht{a}\bar a}\lambda^b_{\pht{b}\bar b}~\ud\phi^{\bar a}
\wedge\ud\phi^{\bar b} \wedge e^c \wedge e^d~\epsilon_{abcd}$ as the volume
form for various different $M^2\times R^2$, where $M^2$ denotes 
the $2$-dimensional submanifolds spanned by an arbitrary pair of $e^a$ and $e^b$.
A linear combination of all the four terms in \eqref {eq:mixing_terms} -- that is a most 
general potential  for GR that includes the CC term -- can be thought as linear 
combination of  all  possible departures of the volume forms
of $\mathcal M$  from those of $\mathcal M^{4-n}\times R^n $, with $n=1,2,3,4$ 
denoting  the number of dimensions that have been ``straightened out''.

The principles outlined here allows one to consider various generalizations.
For example, the dimensionality of $E$ does not have to coincide with the dimensionality of $\mathcal M$.
If the dimensionality of $E$ is $D$, the index $\bar a$ takes values from $1$ to  
$D$ in the vector representation
of $SO(D-1, 1)$, while  the auxiliary fields $\lambda^a_{\pht{a}\bar b}$  transforms as 
bi-vector of $SO(3,1)_{\rm LLT}$ and $SO(D-1, 1)_{\rm INT}$ respectively.  If $D>4$, the extra coordinates will correspond to extra physical scalar fields with a Galileon-like symmetry.  The construction
 remains consistent, in the sense that a Boulware-Deser like ghost will not be introduced.  Such an extension of dRGT was already
considered in \cite{Gabadadze:2012tr}. Its vierbein formulation was given in \cite{Andrews:2013ora} and was used to show ghost-freedom.  The present formalism 
provides the LLT invariant vierbein formulation of this theory.

In the extreme case where $D=1$, $\phi^{\bar a}$ reduces to a single scalar $\phi$ (the index
$\bar a$ takes only one value) and $\lambda^a_{\pht{a}\bar b}$ reduces to a single
Lorentz vector $v^a(x)$ subjected to the condition $v^2=1$. 
Following the discussion given above, one finds that one of the 
natural interaction terms to consider is
\be
L\sim v^a~\ud\phi\wedge e^b\wedge e^c\wedge e^d ~\epsilon_{abcd},
\ee
which, after integrating out $v^a$, gives rise to an action of the Cuscuton type \cite{Afshordi:2006ad}
\begin{equation}
\label{eq:sqrt_scalar}
\mathcal{L}\sim \sqrt{-g}\sqrt{\left|g^{\mu\nu}\partial_\mu\phi\partial_\nu\phi\right|}\,.
\end{equation}

Last but not least, one may consider an even more general class of theories where the internal
global symmetry does not have to be the Lorentz symmetry but is instead described by an arbitrary Lie group $G$. As long as
$\phi^{\bar a}$ is in some representation $R$ of $G$ and the field
$\lambda^a_{\pht{a}\bar b}$ is in the bi-representation of
$R$ and the Lorentz group, we may consider interactions of $\phi^{\bar a}$
with gravity described by the Lagrangians given in \eqref{eq:mixing_terms}.
If one further gauges this internal symmetry, one arrives at a broader
class of theories, which includes the bi-gravity theories considered in \cite{Hassan:2011zd}.

\section{4. The decoupling limit in the first order formulation}

In this Section, we will illustrate the advantages of the first-order formalism for the analysis of the decoupling limit (DL) of massive GR. In addition to reproducing very easily the already well known scalar-tensor interactions that arise in this limit, we will derive an all-orders expression for the DL interactions involving the vector helicity of the massive graviton.  To the best of our knowledge, the vector interactions have previously been unknown in closed form, though partial results are 
available \cite{deRham:2010gu,Koyama:2011wx,Tasinato:2012ze}.

We start by decomposing the vierbein field as before
\beq
e^{~a}_\mu=(\exp{\hat B}) ^a_{~~b}~\hat{e}^{~b}_\mu~,
\eeq
where $\hat B^a_{~b}\equiv B^a_{~b}/\mpl^{1/2}$ is an antisymmetric generator of LLT, $\hat B_{ab}=\eta_{ac}\hat B^{c}_{~b}=-\hat B_{ba}$, while $\hat e$ is the vierbein, symmetric on its lower indices, $\hat e_{\mu \nu}\equiv\hat e^{~b}_\mu\eta_{b\nu }=\hat e_{\nu \mu}.$  The symmetric vierbein and the auxiliary scalars are decomposed into background values and their perturbations as 
\beq
\hat e^{~a}_\mu=\delta^a_{\mu}+\frac{S^{~a}_\mu}{\mpl}, \quad \phi^{\bar a}=\delta^{\bar a}_\mu x^\mu-\pi^{\bar a},
\eeq
where $\pi^{\bar a}=\eta^{\bar a\mu}(\p_\mu\pi/\Lambda^3_3+m A_{\mu}/\Lambda^3_3)$ and $\Lambda_3\equiv\(\mpl m^2\)^{1/3}$. The scalings for various perturbation fields have been chosen so as to recover the correct 
quadratic terms in the decoupling limit of the theory.  In the ghost-free theories at hand, this limit is $m\to 0$, $\mpl\to\infty$, with $\Lambda_3$ held finite \cite{ArkaniHamed:2002sp,Creminelli:2005qk,deRham:2010ik}.

Concentrating first on the $S$ - $\pi$ interactions that result from the Lagrangian \eqref{l2}, one can easily 
see that only terms with a single $S$ and a certain  number of $\pi$'s survive in the decoupling limit
\beq
\mathcal{L}_2^{d.l.}\sim S^{~a}_\mu \(\varepsilon^{\mu\nu \bullet\bullet }\varepsilon_{ab\bullet\bullet}\p^b\p_\nu\pi+\frac{1}{\Lambda^3_3}\varepsilon^{\mu\nu \alpha \bullet}\varepsilon_{abc\bullet}\p^b\p_\nu\pi\p^c\p_\alpha\pi\),\nn
\eeq
where the indices on the $\varepsilon$ symbols are contracted with the help of the unit vierbein.
At linear order, the vierbein and metric perturbations are related as $2 S^{~a}_\mu \eta_{a\nu}=h_{\mu\nu}$, therefore the above scalar-tensor interactions
are nothing but the well-known ghost-free DL interactions of the helicity - 0 and helicity - 2 gravitons in massive GR
\cite {deRham:2010ik}. Including the independent interactions $\mathcal{L}_{3,4}$ with three and four powers of $k$, one equally easily reproduces the remaining $h(\p^2\pi)^3$ interaction of the decoupling limit of massive GR. 

As a next step, we use the above formalism to derive a closed-form expression for the vector-scalar interactions in the DL. To illustrate, we will  start with the case when the two free parameters of dRGT are chosen
so that all the scalar-tensor nonlinear interaction at the scale $\Lambda_3$ identically 
vanish \cite {deRham:2010ik}. For this  parameter choice a linear combination of ${\cal L}_2,  {\cal L}_3$ and 
 ${\cal L}_4$  can be expressed, up to a total derivative, in terms of   ${\cal L}_1$ and a CC term with a tuned
value \cite{Hassan:2011vm}; the resulting theory was dubbed   ``the minimal model".  In  the GCT and LLT 
invariant vierbein  formalism the minimal model takes the form:
\beq
\label{mm}
d^4x \mathcal{L}_{min}=\mpl^2 m^2 \varepsilon_{abcd}~ \( e^a\wedge e^b\wedge e^c\wedge e^d-
4~e^a\wedge e^b\wedge e^c\wedge k^d\)~,
\eeq 
where the one form $k$ is defined in the usual way $k^d =  dx^\beta k^{~d}_\beta$ using the ``k-vierbein" $k^{~a}_\mu\equiv e^{~a}_\mu-\lambda^a_{~\bar a}\p_\mu\phi^{\bar a}$.
In spite of the absence of the nonlinear  helicity-0  interactions with helicity-2  
at the scale $\Lambda_3$, the minimal model has nonlinear interaction terms 
of the vector mode with the helicity-0  at the scale\footnote{It also has  
vector-scalar-tensor interaction terms at  higher scales, such as $\Lambda_2=(\mpl m)^{1/2}$ and/or at  
scales formed by  products of the $\Lambda^2_2$ and $\Lambda^3_3$ scales. These nonlinear 
terms,  up and including  quartic order,  were calculated by L. Berezhiani and G. Chkareuli in 
Spring 2012 (unpublished).} $\Lambda_3$ .

As can be straightforwardly checked, the potentially diverging contributions, \textit{e.g.} of the form $\varepsilon\varepsilon B\p^2\pi$, in fact vanish due to the symmetry properties of the $B$ field (this is precisely what allows to consistently set the scaling of the field $B$ to be $(\mpl)^{-1/2}$). Keeping all finite terms involving $B$ in the decoupling limit and expanding the wedge product in \eqref{mm}, one obtains\footnote{For the sake of simplicity, 
 we will not make distinction between the Lorentzian and spacetime indices of the $B$ - field in the decoupling limit,
 since both are contracted with the flat metric.}
\begin{align}
\mathcal{L}^{d.l.}_{min}\supset 12 \bigg(\Lambda^3_3 B^{\mn} B_{\mn}-B^{\mu\alpha} B_\alpha^{~\nu} (\p_\mu\p_\nu\pi-\eta_{\mn}\Box\pi)-2\Lambda^{3/2}_3B^{\mn}\p_\mu A_\nu\bigg)~.
\end{align}

This is the simplest all-orders expression.  It involves the auxiliary field $B$.  We may, if we like, integrate it out to obtain an expression involving only the physical fields $\pi$ and $A$, at the cost of generating an infinite number of terms.  
In matrix notation (all indices are understood to be contracted with the help of the flat metric), the equation of motion for $B$ yields,
\beq
P_{\mu \nu}^{\alpha\beta} (\pi ) B_{\alpha \beta} = F_{\mu\nu}\,,
\eeq
where $F_{\mn}=\p_\mu A_\nu-\p_\nu A_\mu$ denotes the field strength for the vector mode, and 
$P$ is a tensor of the  schematic form $(\eta \eta + \eta \partial\partial\pi)$ appropriately 
antisymmetrized\footnote{We thank the authors of \cite{Andrew} for pointing out 
a sloppy treatment of $P^{-1}$  in version 1 of this work.    Explicit expressions 
are obtained in \cite{Andrew}.}.  When substituted back into the action, the last equation 
gives the closed-form expression for the vector-scalar interactions in the decoupling limit of the ``minimal" massive GR
\beq
\mathcal{L}^{d.l.}_{min}\supset 6 ~ \text{Tr} ~\bigg[ P^{-1} \cdot F\cdot \partial A \bigg ]~.
\eeq
The lowest - order term in the expansion of the latter Lagrangian in powers of $\p\p\pi$ yields the (correct-sign) kinetic term for the vector, while higher order terms give its interactions with the scalar helicity.

Moving away from the minimal model, for the most general form of the potential the Lagrangian has the following schematic form in the decoupling limit\footnote{We discard explicit vector-scalar interactions of the form $\p A\p^2\pi,~\p A\p^2\pi\p^2\pi, \p A\p A\p^2\pi$ because these turn out to be total derivatives.}
\begin{align}
\mathcal{L}^{d.l.}\sim \Lambda^4_3~\bigg[\frac{B B}{\Lambda_3} \(1+\frac{\p^2\pi}{\Lambda^3_3}+\frac{(\p^2\pi)^2}{\Lambda^6_3}+\frac{(\p^2\pi)^3}{\Lambda^9_3} \) 
  +\frac{B \p A}{\Lambda_3^{5/2}}\(1+\frac{\p^2\pi}{\Lambda^3_3}+\frac{(\p^2\pi)^2}{\Lambda^6_3}\)   \bigg]~.
\end{align}
Varying w.r.t. the non-dynamical field $B$  yields an expression for it 
in terms of $\pi$ and $A$, that can be substituted back into the action, recovering the complete 
decoupling limit form of the vector-scalar interactions.  These interactions are derived 
in \cite {Andrew}. The resulting expressions can be readily used for studying dynamics of 
the given sector of the theory on various background solutions.

\section{5. Dynamical antisymmetric field}

While in pure massive gravity the link fields $\lambda^a_{~\bar a}$ are non-dynamical, one can go further and consider a generalization with dynamical link fields, nonlinearly realizing the symmetry breaking pattern \eqref{sb}.  Given the symmetries at hand, the most general Lagrangian at low energy can be written as a function of the fields with definite transformation properties under GCT$\times$ LLT$\times$ $ISO(3,1)_{\text{INT}}$~,
\beq
S=\int d^4x\,\mathcal{L}\(\lambda^{a}_{~\bar a},\phi^{\bar a},e_\mu^{~a},D_{\mu}\)~.
\eeq
The covariant derivative $D_\mu$ acts on the LLT indices through the standard expression $D_\mu\lambda^a_{~\bar a}=\p_\mu\lambda^a_{~\bar a}+\omega_{\mu~~b}^{~a}\lambda^b_{~\bar a}$, where the spin connection $\omega_{\mu~~b}^{~a}$ can be expressed in terms of the vierbein and its derivatives in a torsion-free theory. Being a Lorentz matrix-valued field, $\lambda$ is most conveniently expressed in terms of the antisymmetric generator, $\lambda=\exp(v/f)$, where $f$ denotes the ``decay constant" of $v$.  The decay constant $f$ is an adjustable parameter of the theory.

The lowest-order non-trivial invariant that one can form from these fields can be written as follows
\beq
\mathcal{L}=-f^2\(D_\mu\lambda\)^2=-\eta_{ab}\eta^{\bar a\bar b}\p_\mu v^a_{~\bar a}\p^\mu v^b_{~\bar b}+\dots~\nn,
\eeq
and includes the kinetic term for the six degrees of freedom present in $v^a_{~\bar a}$.  Note that the kinetic term for the $\lambda$ field, alongside with the EH term, 
breaks the local $SL(4)$ symmetry  of the mass and potential terms if we were to promote 
the $\lambda$  to a $SL(4)$ valued field.  We will write $f\sim  \hat f (\mpl\Lambda_3)^{1/2}$, where $\hat f$ is dimensionless.
The $\Lambda_3$ decoupling limit 
remains intact as long as $\hat f$ remains fixed in this limit, i.e. does not depend parametrically on any other scales.

Supplementing the action by the ghost-free potential terms, for example Eq. \eqref{l2}, one obtains a set of  interactions of $v$ with the rest of the fields present in the theory (for the moment, we choose the LLT gauge defined by $B=0$.) At the linearized order, \eqref{l2} yields the mass term, as well as a mixing with the vector mode in the decoupling limit (we disregard the distinction between the LLT and spacetime indices for notational simplicity) 
\beq
\mathcal{L}^{d.l.}=v^{\mn}(\Box+{2\Lambda^2_3\over \hat f^2}) v_{\mn}+{2\Lambda_3\over \hat f} F^{\mn} v_{\mn}~.
\eeq
A shift in the $v$ field, $v_{\mn}\to \hat v_{\mn}-\frac{F_{\mn}}{{\Lambda_3\over \hat f } \(2+\frac{\hat f^2\Box}{\Lambda^2_3}\)}$
diagonalizes the action, bringing it to the following form
\beq
\mathcal{L}^{d.l.}=\hat v^{\mn}\(\Box+{2\Lambda^2_3\over \hat f^2}\) \hat  v_{\mn}-F^{\mn} \frac{1}{2+\frac{\hat f^2\Box}{\Lambda^2_3}} F_{\mn}.
\eeq
Another peculiar feature of the above action is that $\hat v$ acquires a mass $|m_v^2|\sim\Lambda^2_3/{\hat f}^2$.  This is below the cutoff of the effective theory to the extent that $\hat f\gg 1$. Note that, in order to reproduce the correct sign of the vector kinetic term at low energies, $m^2_v$ 
has to be tachyonic; however, one could expect higher powers of $v$ (e.g. $v^4$) to also be present, and these could stabilize the $v$ potential.  Likewise, the kinetic term of the vector acquires a modification. 
In the regime,  $\hat f^2\Box/\Lambda^2_3\ll 1$, the modification is irrelevant and $A_{\mu}$  propagates the usual two vector polarizations of the massive graviton.  Note  that the residue of the vector 
particle propagator vanishes at the position of the pole of the $v$ field.

One can give the above generation of the mass $m_v$ an Anderson mechanism - like interpretation. Indeed, both of the antisymmetric fields, $B$ and $v$, nonlinearly realize the local Lorentz invariance. One can always choose a gauge in which either of the two, e.g. $B$, is frozen to be zero, however one combination of these is gauge invariant (at the linear level, the invariant combination is simply $\hat f\Lambda_3^{1/2} B-v$).  Then, the gauge-invariant combination (which reduces to $v$ in the $B=0$ gauge) acquires a mass due to the spontaneous breaking of LLT.

Finally, we comment on ghost-freedom of the interactions of the antisymmetric field $v$ with the rest of the modes, present in the decoupling limit Lagrangian. The object $k_a^{~\mu}$ is decomposed (excluding the symmetric vierbein perturbation) in the $B=0$ gauge as follows,
\ba
k_a^{~\mu}&=&\frac{\p_\mu\p^a\pi}{\Lambda^3_3}+\frac{\p_\mu A^a}{\mpl^{1/2} \Lambda^{3/2}_3}-\frac{v^a_{~\mu}}{\hat f (\mpl\Lambda_3)^{1/2}}+\frac{v^a_{~ b} \p_\mu\p^b\pi}{\hat f\mpl^{1/2}\Lambda_3^{7/2}}
+\frac{•v^a_{~b}\p_\mu A^b}{•\hat f \mpl\Lambda^2_3}-\frac{v^a_{~b} v^b_{~\mu}}{2\hat f^2\mpl\Lambda_3}+\frac{v^a_{~b} v^b_{~c}\p_\mu\p^c\pi}{•2\hat f^2\mpl\Lambda_3^4}+\dots\nn
\ea
Most of the terms, that follow from the expansion of \eqref{l2} are easily checked to be safe from more that two derivatives acting on fields in the resulting equations of motion -- either on the basis of 
antisymmetry of $v$, or due to the presence of the $\varepsilon$ symbols in the corresponding expressions.

The only two interactions for which this property is not apparent are of the $v v \p\p\pi\p\p\pi$-type. The first of these is $\varepsilon^{\mn\bullet\bullet}\varepsilon_{ab\bullet\bullet}v^a_{~\rho}\p_\mu\p^\rho\pi v^b_{~\sigma}\p_\nu\p^\sigma\pi$. The only potentially dangerous, three-derivative term arises in the equation of motion for $\pi$ (all other similar terms vanish by antisymmetrization), and has the following form, $$\varepsilon^{\mn\bullet\bullet}\varepsilon_{ab\bullet\bullet}\p_\mu (v^a_{~\rho}v^b_{~\sigma})\p^\rho \p^\sigma\p_\nu\pi~.$$
Now, antisymmetrization in the $a$ and $b$ indices tells us that the object in the parentheses is antisymmetric in the $(\rho,\sigma)$ pair. Contracted with $\p^\rho\p^\sigma$ on the scalar, the term at hand vanishes. Likewise, a potentially dangerous term in the Lagrangian $\varepsilon^{\mn\bullet\bullet}\varepsilon_{ab\bullet\bullet}\p_\mu\p^a \pi v^b_{~\rho} v^{\rho}_{~\sigma}\p_\nu\p^\sigma\pi$ yields an apparently ghostly contribution to the $\pi$-equation of motion
\beq
\varepsilon^{\mn\bullet\bullet}\varepsilon_{ab\bullet\bullet}\big [\p_\mu (v^b_{~\rho} v^\rho_{~\sigma})\p^a\p_\nu\p^\sigma\pi+\p_\nu (v^b_{~\rho} v^\rho_{~\sigma})\p^a\p_\mu\p^\sigma\pi\big ]~.\nn
\eeq
However, the object in the square parentheses in this expression is manifestly symmetric under $\mu\to\nu$. Contracted with the antisymmetric $\varepsilon^{\mn\bullet\bullet}$, this again yields zero. 
Of course, although this is a nice consistency-check, such a vanishing of the three-derivative terms in the equations of motion is by no means surprising and follows automatically from the inherent ghost-freedom of the potential \eqref{l2}.

\section{6. Brief comments on the literature\label{comments}}

In this section, we briefly discuss the status of massive gravity as applied to the real world.
In this approach, the graviton mass is taken to  be of the order of  the present day Hubble 
parameter, $m\sim H_0\sim 10^{-33}~{\rm eV}$  (for phenomenological bounds on the
graviton mass see \cite {Goldhaber:2008xy}).  Although this is a very  small parameter 
as compared to the Planck scale, such smallness is robust -- the mass parameter does 
not get renormalized by large quantum corrections \cite {deRham:2012ew,ArkaniHamed:2002sp}; this is unlike
the cosmological constant which does receive large renormalizations.  Therefore, it is appealing 
to describe the observed cosmic acceleration as an effect due to a nonzero graviton mass. 

Massive gravitons  can produce a state with the stress-tensor mimicking dark energy 
(the so--called self-accelerated solutions \cite{deRham:2010tw,Koyama:2011xz,D'Amico:2011jj,Volkov:2012cf,Gratia:2012wt,Gumrukcuoglu:2011ew}).  Massive gravity dark energy is expected to 
have a slightly different predictions from those of CC based cosmology,  and the differences 
may be tested observationally. 
These solutions produce dark energy with the equations of state identical to that of CC, but different 
fluctuations.  Unfortunately,  certain  fluctuations  about these solutions are problematic -- 
some of the physical 5 degrees of freedom 
have vanishing kinetic terms,  destabilizing  the background 
\cite {Gumrukcuoglu:2011zh}.  Extensions of dRGT by  additional scalars 
\cite {D'Amico:2011jj,D'Amico:2012zv} or bi and multi-gravity \cite {Hassan:2011zd,Hinterbichler:2012cn},
or further extensions \cite {Comelli,QG}, also exhibit self-accelerated solutions.  Recently, an extension 
by scalars has been proposed  by De Felice and Mukohyama \cite {DeFelice:2013tsa} and shown 
 to have a  self-accelerated solution with 
stable fluctuations --  a first example of  this  kind.

Spherically symmetric solutions and black holes in massive GR have been studied in \cite{Koyama:2011xz,sphsol}.
A general issue in dRGT is that it is a strongly coupled theory at the distance scale $(\Lambda_3)^{-1}$, 
which for the above value of the graviton mass is $\sim 1000$ km. 
This scale is background dependent, and  decreases   for realistic backgrounds 
\cite {Berezhiani:2013dca}, but never enough for one to feel comfortable with it. 
The higher dimensional operators -- that best manifest themselves in the decoupling limit --
are suppressed by this scale.  Moreover, on realistic backgrounds these operators 
give rise to order 1 or larger classical renormalization of the kinetic terms 
of fluctuations. That is how some of these kinetic terms vanish
or flip their signs  on the self-accelerated backgrounds.  Therefore, dRGT needs an 
extension beyond the strong coupling scale in order for it to be potentially 
applicable to the real world. This extension is unknown at present,  but for it to work it should introduce new
states at or below the scale $\Lambda_3$. Therefore, many properties of the backgrounds and 
fluctuations sensitive to scales above $\Lambda_3$ can get modified in an extended theory\footnote{For related recent developments see Refs. \cite {Denis}.}. 

Furthermore, in the decoupling limit dRGT  gets related 
\cite {deRham:2010ik} to the Galileons \cite {Nicolis:2008in}. 
The latter are known to exhibit superluminal propagation on nontrivial backgrounds.
So does dRGT for a large portion of the $\alpha_3,\alpha_4$ parameter 
space.  For theories satisfying the Froissart  bound, this has been argued \cite{Adams:2006sv} to preclude
a standard UV completion by a local, Lorentz invariant field or string theory, 
however,  theories with long-rage fields  do not necessarily obey this bound;  moreover, 
there is no claim to rule out a possible Lorentz-violating, non-local or intrinsically higher
dimensional completion.  Furthermore, there  is an  exception  for some special values 
of $\alpha_3,\alpha_4$, where subluminality  for a spherically symmetric solution 
is achieved at the expense of not having an asymptotically flat background\footnote{Although not directly related to massive gravity, cosmological solutions with subluminal spectra in dilatation invariant theories of Galileons have been found in \cite{Creminelli:2012my,Hinterbichler:2012fr,Hinterbichler:2012yn}.} \cite {Berezhiani:2013dw,Berezhiani:2013dca}.  

The question of whether superluminality  can lead to prohibitive  
acausality is entangled with the strong coupling issue  
\cite {Burrage:2011cr}. The conclusion of acausality of massive gravity 
\cite {Deser:2012qx,Deser:2013eua}  that  has been reached 
by constructing superluminal  shock waves and characteristics
is,  in the 
context of a low energy  theory, not warranted without a further  nuanced study.    A well known counterexample is the following: quantum electrodynamics 
(QED) in an external gravitational field,  at energies below the  electron mass,  
gives rise to dimension  6 operators, one of which  yields  superluminal 
characteristics for a photon propagating in a given non-trivial gravitational background
\cite {Drummond:1979pp}. However, this superluminality -- which appears within the effective 
theory -- does not mean that QED supplemented by GR is an acausal theory. 
In spite of a large body of literature on the issue of superluminality vs. acausality, 
some with split views, we believe that the low energy effective field theory understanding of systematic criteria for 
potential harms, or their absence,  of superluminal low energy group and phase velocities 
is still to be precisely formulated \cite {WorkInProgress}.

\vspace{10mm}

{\it Notes added:}  Ref. \cite {Andrew} has studied the decoupling limit of dRGT 
using the vierbein formalism. This work, even though it appeared later than v1
of the present work, should be considered as concurrent on the main idea
of studying the decoupling limit in this formalism; moreover  the results 
of \cite{Andrew} on the decoupling limit 
are superior to ours in their completeness. 

The remarkable work \cite {Peloso}, appearing in 2006, introduced almost all of the ingredients of massive gravity, 
including the St\"uckelbergs for the LLT's (but not the $\phi^a$ fields).  Unfortunately,  Ref. \cite {Peloso}
adopts an incorrect conclusion regarding the existence of the Boulware-Deser ghost.  We thank 
Andrew Tolley for bringing this to our attention.

\vspace{10mm}

{\bf Acknowledgments:}
The authors benefited from communications with
Nikita Nekrasov, Warren Siegel, Andrew Tolley, and Arkady Vainshtein, on topics related to the subject of 
the present  work.  GG is supported by the NSF grant
PHY-0758032 and NASA grant NNX12AF86G S06.
Research at Perimeter Institute is supported by the Government of Canada through Industry Canada 
and by the Province of Ontario through the Ministry of Economic Development and Innovation. The 
work of KH was made possible in part through the support of a grant from the John Templeton 
Foundation.  The work of YS is supported by the John Templeton Foundation through Professor John Moffat.
The opinions expressed in this publication are those of the authors and do not necessarily 
reflect the views of the John Templeton Foundation.  DP has been supported by the U.S. Department of 
Energy under contract No. DE-SC0009919.  GG would like to thank the Perimeter Institute for hospitality.  KH and DP would like to thank the Center for Particle Physics and 
Cosmology at New York University for their hospitality.   


\end{document}